\documentclass[12pt]{article}

\usepackage{bbm,latexsym}
\usepackage{amsmath}
\usepackage{amsfonts}
\usepackage{epsfig}

\newcommand{\be}{\begin{equation}}
\newcommand{\ee}{\end{equation}}
\newcommand{\ba}{\begin{eqnarray}}
\newcommand{\ea}{\end{eqnarray}}
\newcommand{\no}{\nonumber\\}

\newcommand{\gtrsim}{\:\mbox{\raisebox{-3pt}{$\stackrel%
{\displaystyle >}{\sim}$}}\:}

\newcommand{\mnu}{\mathcal{M}_\nu}

\renewcommand{\aa}{\left| a_4 \right|}
\newcommand{\mm}{\left| \mu_\mathrm{m} \right|}
\newcommand{\zz}{\mathbbm{Z}_2}

\begin{document}

\title{\normalsize \hfill UWThPh-2008-7 \\[1cm]
\LARGE
A light pseudoscalar in a model \\
with lepton family symmetry $O(2)$
}
\setcounter{footnote}{2}
\author{Walter Grimus,$^{(1)}$\thanks{E-mail: walter.grimus@univie.ac.at} 
\
Lu\'\i s Lavoura,$^{(2)}$\thanks{E-mail: balio@cftp.ist.utl.pt}
\
and David Neubauer$^{(1)}$\thanks{E-mail: david\_neubauer@hotmail.com}
\\[4mm]
$^{(1)} \!$ \small 
Faculty of Physics, University of Vienna \\[-1mm]
\small Boltzmanngasse 5, A--1090 Vienna, Austria
\\*[2mm]
$^{(2)} \!$ \small
Universidade T\'ecnica de Lisboa
and Centro de F\'\i sica Te\'orica de Part\'\i culas \\[-1mm]
\small
Instituto Superior T\'ecnico, 1049-001 Lisboa, Portugal\\*[6mm]}

\date{23 June 2008}

\maketitle

\begin{abstract}
We discuss a realization of the non-abelian group $O(2)$
as a family symmetry for the lepton sector.
The reflection contained in $O(2)$
acts as a $\mu$--$\tau$ interchange symmetry, 
enforcing---at tree level---maximal atmospheric neutrino mixing
and a vanishing mixing angle $\theta_{13}$. 
The small ratio $m_\mu/m_\tau$ (muon over tau mass)
gives rise to a suppression factor
in the mass of one of the pseudoscalars of the model.
We argue that such a light pseudoscalar
does not violate any experimental constraint.
\end{abstract}

\newpage

\section{Introduction}
\label{intro}

The discoveries of solar- and atmospheric-neutrino oscillations,
besides having constituted remarkable experimental feats
and having given neutrino theorists a much needed shot in the arm,
brought with them the pleasant surprise that
two of the lepton mixing angles 
seem to have 
(or are, at least, not far from)
extreme values.
Indeed,
contrary to the solar-neutrino mixing angle,
which has a large but non-maximal value,
the atmospheric-neutrino mixing angle could be maximal ($\pi / 4$)
and the third mixing angle,
$\theta_{13}$,
might vanish.
These two features are easily explained,
theoretically,
by assuming the (effective) light-neutrino Majorana mass matrix $\mnu$,
in the weak basis where the charged-lepton mass matrix is diagonal,
to be $\mu$--$\tau$ symmetric~\cite{mu-tau,HS,mu-tau-recent}:
\be
\label{Mnu}
\mnu = \left( \begin{array}{ccc}
x & y & y \\ y & z & w \\ y & w & z
\end{array} \right).
\ee
The mass matrix~(\ref{Mnu}) is in very good agreement
with the presently known data~\cite{fits}.
Let us write the $\mu$--$\tau$ interchange symmetry as
\be
\label{s}
s: \
D_{\mu L} \leftrightarrow D_{\tau L},
\
\mu_R \leftrightarrow \tau_R,
\
\nu_{\mu R} \leftrightarrow \nu_{\tau R},
\
\phi_1 \leftrightarrow \phi_2,
\ee
where the $D_{\alpha L}$ ($\alpha = e, \mu, \tau$)
are the left-handed-lepton gauge-$SU(2)$ doublets,
the $\nu_{\alpha R}$ are right-handed-neutrino $SU(2)$ singlets,
which we add to the theory
in order to enable a seesaw mechanism~\cite{seesaw}, 
and the $\phi_j$ ($j = 0, 1, 2$)
are three Higgs doublets.
This $\mu$--$\tau$ interchange symmetry $s$ allows one to relate
the small ratio of muon mass over tau mass,
$m_\mu / m_\tau$,
to a small ratio of vacuum expectation values (VEVs).\footnote{A
similar mechanism has previously been used,
for instance, in~\cite{ma}. There, 
the ratio between the up- and charm-quark masses
is equal to a ratio of two VEVs, in a model with
horizontal symmetry $S_3 \times \mathbbm{Z}_3$.}
Indeed,
if there is in the theory
some extra family symmetry---besides $s$---such that
only $\phi_1$ has Yukawa couplings to the muon family
and only $\phi_2$ has Yukawa couplings to the tau family,
then
\be
\mathcal{L}_Y = \cdots - y_4 \left(
\bar D_{\mu L} \phi_1 \mu_R +
\bar D_{\tau L} \phi_2 \tau_R \right)
+ \mbox{H.c.},
\ee
hence $m_\mu / m_\tau = \left| v_1 / v_2 \right|$,
where $v_j \left/ \sqrt{2} \right. = 
\left\langle 0 \left| \phi_j^0 \right| 0 \right\rangle$
is the vacuum expectation value (VEV)
of the neutral component of $\phi_j$.
This may allow one to relate a property of the charged-lepton spectrum
to features of the scalar potential and spectrum.

In this paper we present a model in which the small ratio $m_\mu / m_\tau$
is related to a suppression factor in the mass of one of the pseudoscalars.
One thus has an indirect connection between neutrino mixing properties
and features of the scalar sector.
Our model is particularly simple
in that it uses the non-abelian group $O(2)$
as its main family symmetry.
It has a scalar potential with less parameters than previous models,
predicting in particular no $CP$ violation.

In section~\ref{model}
we present the symmetries and the Lagrangian of our model.
In section~\ref{scalar sector}
we study the mass matrices of the scalars.
Section~\ref{phenomenology}
is devoted to experimental constraints on our model.
We summarize our findings in section~\ref{concl}.
Three appendices contain material
which may be omitted in a first reading of our paper.
Appendix~A makes an abstract description of the group $O(2)$.
Appendix~B compares the present model
with a previous model of maximal atmospheric-neutrino mixing
with a naturally suppressed ratio $m_\mu / m_\tau$~\cite{z2,smallratio}.
Appendix~C presents a variation of our model
in which the symmetry $s$ is substituted
by a non-standard $CP$ symmetry~\cite{CP},
with the practical consequence that one predicts
``maximal $CP$ violation'' in lepton mixing
instead of a vanishing mixing angle $\theta_{13}$.

\section{The model}
\label{model}

We consider an extension of the standard electroweak model (SM)
with gauge group $SU(2) \times U(1)$, with multiplets as described in
the introduction:
left-handed $SU(2)$ doublets $D_{\alpha L}$, 
right-handed $SU(2)$ singlets $\alpha_R$ and $\nu_{\alpha R}$
($\alpha = e,\, \mu,\, \tau$),
and three Higgs doublets $\phi_j$ ($j=0,1,2$).

The family symmetries of our model
are the reflection symmetry $s$ in~(\ref{s})
and also a $U(1)$ symmetry acting on the multiplets as 
\be
\label{U1}
U(1): \
\left\{ 
\begin{array}{rcl}
\left( D_{\mu L}, \, \tau_R, \, \nu_{\mu R} \right)
&\to& 
e^{+i \theta} \left( D_{\mu L}, \, \tau_R, \, \nu_{\mu R} \right),
\\ 
\left( D_{\tau L}, \, \mu_R, \, \nu_{\tau R} \right)
&\to&
e^{-i \theta} \left( D_{\tau L}, \, \mu_R, \, \nu_{\tau R} \right),
\\
\phi_1 &\to& e^{+2 i \theta}\, \phi_1,
\\
\phi_2 &\to& e^{-2i\theta}\, \phi_2.
\end{array}
\right.
\ee
Moreover,
we need an extra $\zz$ symmetry
(beyond $s$)
given by
\be
\label{Z2}
\zz:
\
\nu_{e R}, \
\nu_{\mu R}, \
\nu_{\tau R}, \
e_R, \
\phi_0 \
\mbox{change sign}.
\ee
The symmetry $U(1)$ in~(\ref{U1})
does not commute with the symmetry $s$ in~(\ref{s}).
One can conceive $U(1)$ and $s$ as generating together
the non-abelian group $O(2)$,
as discussed in appendix~A.
That appendix also contains the irreducible representations of $O(2)$.
Equations~(\ref{s}) and~(\ref{U1}) may be interpreted
in terms of those irreducible representation
by the following assignments
\be
\label{f}
\begin{array}{rcl}
\underline{1}: & & D_{e L}, \ \nu_{eR}, \ e_R, \ \phi_0;
\\
\underline{2}^{(1)}: & &  \left( D_{\mu L}, \, D_{\tau L} \right), \
\left( \tau_{R}, \, \mu_{R} \right), \
\left( \nu_{\mu R}, \, \nu_{\tau R} \right);
\\
\underline{2}^{(2)}: & &
\left( \phi_{1}, \, \phi_{2} \right).
\end{array}\ee
The full family symmetry of the model is thus $G = O(2) \times \zz$.

The above multiplets and symmetries determine the Yukawa Lagrangian 
\ba
\mathcal{L}_Y & = &
- y_1\, \bar D_{eL} \tilde \phi_0 \nu_{eR}
- y_2 \left( \bar D_{\mu L} \tilde \phi_0 \nu_{\mu  R}
+ \bar D_{\tau L} \tilde \phi_0 \nu_{\tau R} \right)
\no &&
- y_3\, \bar D_{e L} \phi_0 e_R
- y_4 \left( \bar D_{\mu  L} \phi_1 \mu_R 
+ \bar D_{\tau  L} \phi_2 \tau_R \right)
+ \mbox{H.c.,}
\label{LY}
\ea
where $\tilde \phi_j \equiv i \tau_2 \phi_j^\ast$.
Because of the $\zz$ symmetry of~(\ref{Z2})
only $\phi_0$ couples to the $\nu_{\alpha R}$ and to $e_R$.
Because of the $U(1)$ symmetry of~(\ref{U1})
the Yukawa-coupling matrices are all diagonal.
Due to the $\mu$--$\tau$ interchange symmetry of~(\ref{s}) 
the neutrino Dirac mass matrix is given by
\be
\label{MD}
M_D = \mbox{diag} \left(a, \, b, \, b \right),
\ee
with $a = y_1^\ast v_0 \left/ \sqrt{2} \right.$
and $b = y_2^\ast v_0 \left/ \sqrt{2} \right.$.  
The charged-lepton masses are
\be
m_e = \frac{\left| y_3 v_0 \right|}{\sqrt{2}}, \quad
m_\mu = \frac{\left| y_4 v_1 \right|}{\sqrt{2}}, \quad
m_\tau = \frac{\left| y_4 v_2 \right|}{\sqrt{2}}.
\ee
There is one VEV per charged-lepton mass.
The mass ratio
\be
\label{ratio}
\frac{m_\mu}{m_\tau} = \left| \frac{v_1}{v_2} \right|
\ee
is determined solely by a ratio of VEVs,
the Yukawa couplings being totally absent therefrom.

An important ingredient of the model
is the soft breaking of the $U(1)$ of~(\ref{U1})---but neither of $s$
nor of $\zz$---by terms in the Lagrangian of dimension three or smaller.
The family symmetry group $O(2)$
is softly broken to $s$:
\be
\label{softbr}
O(2) \times \zz
\ \stackrel{\mathrm{soft}}{\longrightarrow} \ 
\zz^{(s)} \times \zz,
\ee
where
$\zz^{(s)}$ is the $\zz$ group generated by $s$.
Later,
$\zz^{(s)} \times \zz$ is spontaneously broken
by the VEVs of the Higgs doublets.
The soft breaking~(\ref{softbr}) permits
the right-handed neutrino singlets to acquire Majorana mass terms,
\be
\label{LM}
\mathcal{L}_\mathrm{M} = \frac{1}{2}\, \nu_R^T C^{-1} \! M_R^* \nu_R
+ \mbox{H.c.,}
\ee
satisfying $\left( M_R \right)_{e \mu} = \left( M_R \right)_{e \tau}$
and $\left( M_R \right)_{\mu \mu} = \left( M_R \right)_{\tau \tau}$
because of $s$.
Together,
equations~(\ref{MD}) and (\ref{LM})
determine the form of the effective Majorana mass matrix
of the light neutrinos,
$\mnu = - M_D^T M_R^{-1} M_D$,
to be as in equation~(\ref{Mnu}).
As stated in section~\ref{intro},
this form of $\mnu$ leads to two of the three lepton mixing angles
having extreme values:
$\theta_{23} = \pi / 4$ and $\theta_{13} = 0$,
while the remaining mixing angle $\theta_{12}$,
and also the neutrino masses and Majorana phases,
remain undetermined.

\section{The scalar sector}
\label{scalar sector}

\subsection{The scalar potential and its minimum}
\label{potential}

The scalar potential,
taking into account the symmetries of the model,
is given by
\ba
V &=&
\mu_0\, \phi_0^\dagger \phi_0
+ \mu_{12} \left( \phi_1^\dagger \phi_1 + \phi_2^\dagger \phi_2 \right)
+ \mu_\mathrm{m} \left( \phi_1^\dagger \phi_2 + \phi_2^\dagger \phi_1 \right) 
\no
& &
+ a_1 \left( \phi_0^\dagger \phi_0 \right)^{2}
+ a_2\, \phi_0^\dagger \phi_0
\left( \phi_1^\dagger \phi_1 + \phi_2^\dagger \phi_2 \right)
+ a_3 \left( \phi_0^\dagger \phi_{1} \, \phi_1^\dagger \phi_0
+ \phi_0^\dagger \phi_2 \, \phi_2^\dagger \phi_0 \right)
\no
& &
+ a_4\, \phi_0^\dagger \phi_1 \, \phi_0^\dagger \phi_2
+ a_4^\ast\, \phi_1^\dagger \phi_0 \, \phi_2^\dagger \phi_0
+ a_5 \left[ \left( \phi_1^\dagger \phi_1 \right)^2
+ \left( \phi_2^\dagger \phi_2 \right)^2 \right]
\no
& &
+ a_6\, \phi_1^\dagger \phi_1 \, \phi_2^\dagger \phi_2
+ a_7\, \phi_1^\dagger \phi_2 \, \phi_2^\dagger \phi_1.
\label{pot}
\ea
Because of the soft breaking~(\ref{softbr})
we have added to the potential
a term $\phi_1^\dagger \phi_2 + \phi_2^\dagger \phi_1$,
which breaks the $U(1)$ of~(\ref{U1}) but not the $s$ of~(\ref{s}).
In equation~(\ref{pot}) we are implicitly assuming that
there are in the model no scalar multiplets beyond $\phi_{0,1,2}$.
The quarks have Yukawa couplings to $\phi_0$
but neither to $\phi_1$ nor to $\phi_2$---this situation may be enforced
by suitably extending the symmetry $\zz$ of~(\ref{Z2})
to the quark sector.

All the parameters in equation~(\ref{pot}) are real,
except $a_4$ which is in general complex.
There are in the potential only two terms,
the $a_4$ term and the $\mu_\mathrm{m}$ term,
which can feel the two relative phases
among the VEVs of the three doublets.
Therefore,
one can adjust those phases such that,
simultaneously,
the VEVs $v_{0,1,2}$ are real and positive
while both $\mu_\mathrm{m}$ and $a_4$ are real and negative.
This arrangement minimizes $V$.
Thus,
from now on we shall use
\be
\label{transf}
\mu_\mathrm{m} = - \mm, \ a_4 = - \aa, \ v_0 > 0, \ v_1 > 0, \ v_2 > 0.
\ee

The VEVs must fulfill two conditions:
\ba
\label{cond0}
v &\simeq& 246 \ \mbox{GeV}, 
\\
\label{cond1}
\frac{v_1}{v_2} &=& \frac{m_\mu}{m_\tau},
\ea
where
\be
v \equiv \sqrt{v_0^2 + v_1^2 + v_2^2}.
\ee
The condition~(\ref{cond0})
follows from the assumption that
there are in the model no scalar multiplets beyond $\phi_{0,1,2}$.

Writing $\left\langle 0 \left| V \right| 0 \right\rangle$
as a function of $v_0^2$,
$v_1^2 + v_2^2$,
and $v_1 v_2$,
and enforcing the stability
of $\left\langle 0 \left| V \right| 0 \right\rangle$
relative to each of these parameters,
one obtains,
respectively,
\ba
\mu_0 &=& - a_1 v_0^2 - \frac{B}{2} \left( v_1^2 + v_2^2 \right)
+ \aa v_1 v_2,
\\
\mu_{12} &=& - \frac{B}{2}\, v_0^2 - a_5 \left( v_1^2 + v_2^2 \right),
\\
\label{mm}
\mm &=& \frac{A}{2}\, v_1 v_2 - \frac{\aa}{2}\, v_0^2,
\ea
where
\ba
A &\equiv& a_6 + a_7 - 2 a_5,
\\
B &\equiv& a_2 + a_3.
\ea
These equations allow one to replace in a systematic way
the parameters $\mu_0$,
$\mu_{12}$,
and $\mu_\mathrm{m}$ by the VEVs.
This replacement is convenient in order to calculate
the masses of the scalars of the model
in terms of independent parameters.
Notice that it follows from equation~(\ref{mm}) that $A > 0$.

\subsection{The mass matrices of the scalars}

We parameterize the Higgs doublets as
\be
\label{varphi}
\phi_j = \left(
\begin{array}{c} \varphi_j^+ \\
\left( v_j + \rho_j + i \eta_j \right) \left/ \sqrt{2} \right.
\end{array} \right),
\ee
with real fields $\rho_j$ and $\eta_j$.
Since with our convention there are
neither complex couplings in $V$ nor complex VEVs,
$CP$ is conserved in the scalar sector,
the fields $\rho_j$ are scalars
while the $\eta_j$ are pseudoscalars,
and there is no scalar--pseudoscalar mixing.
The mass terms of the scalars are given by
\ba
\mathcal{L}_\mathrm{scalar \ masses} &=& 
- \left( \varphi^-_0, \, \varphi^-_1, \, \varphi^-_2 \right)
\mathcal{M}^2_\varphi
\left( \begin{array}{c}
\varphi^+_0 \\ \varphi^+_1 \\ \varphi^+_2
\end{array} \right)
\no & &
- \frac{1}{2}\,
\left( \rho_0, \, \rho_1, \, \rho_2 \right)
\mathcal{M}^2_\rho
\left( \begin{array}{c} \rho_0 \\ \rho_1 \\ \rho_2 \end{array} \right)
- \frac{1}{2}\,
\left( \eta_0, \, \eta_1, \, \eta_2 \right)
\mathcal{M}^2_\eta
\left( \begin{array}{c} \eta_0 \\ \eta_1 \\ \eta_2 \end{array} \right).
\ea
After some algebra we find that the scalar mass matrices are
\ba
\mathcal{M}^2_\varphi &=&
\frac{a_3}{2} \left( \begin{array}{ccc}
- \left( v_1^2 + v_2^2 \right) & v_0 v_1 & v_0 v_2 \\
v_0 v_1 &  - v_0^2 & 0 \\
v_0 v_2 & 0 & - v_0^2 
\end{array} \right)
+ \frac{\aa}{2} \left( \begin{array}{ccc}
2 v_1 v_2 & - v_0 v_2 & - v_0 v_1 \\
- v_0 v_2 & 0 & v_0^2 \\
- v_0 v_1 & v_0^2 & 0
\end{array} \right)
\no & &
+ \frac{2 a_5 - a_6}{2} \left( \begin{array}{ccc}
0 & 0 & 0 \\
0 & - v_2^2 & v_1 v_2 \\
0 & v_1 v_2 & - v_1^2
\end{array} \right),
\label{M+}
\\
\label{MR}
\mathcal{M}^2_\rho &=&
\frac{1}{2} \left( \begin{array}{ccc}
4 a_1 v_0^2 &
2 v_0 \left( B v_1 - \aa v_2 \right) &
2 v_0 \left( B v_2 - \aa v_1 \right) \\
2 v_0 \left( B v_1 - \aa v_2 \right) &
4 a_5 v_1^2 + A v_2^2 &
\left( 4 a_5 + A \right) v_1 v_2 \\
2 v_0 \left( B v_2 - \aa v_1 \right) &
\left( 4 a_5 + A \right) v_1 v_2 &
4 a_5 v_2^2 + A v_1^2
\end{array} \right),
\\
\label{MI}
\mathcal{M}^2_\eta &=&
\aa \left( \begin{array}{ccc}
2 v_1 v_2 &
- v_0 v_2 &
- v_0 v_1 \\
- v_0 v_2 &
0 & 
v_0^2 \\
- v_0 v_1 &
v_0^2 &
0 
\end{array} \right)
+
\frac{A}{2} \left( \begin{array}{ccc}
0 &
0 &
0 \\
0 &
v_2^2 & 
- v_1 v_2 \\
0 &
- v_1 v_2 & 
v_1^2 
\end{array} \right).
\ea

Both $\mathcal{M}^2_\varphi$ and $\mathcal{M}^2_\eta$
have an eigenvector
\be
\label{X0}
X_0 = \frac{1}{v}
\left(  \begin{array}{c} v_0 \\ v_1 \\ v_2 \end{array} \right)
\ee
with eigenvalue zero;
the corresponding scalar fields
are the unphysical scalars (Goldstone bosons) $G^\pm$ and $G^0$, 
associated with the $W^\pm$ and $Z^0$ gauge bosons,
respectively.
We denote the diagonalization of $\mathcal{M}^2_\varphi$,
$\mathcal{M}^2_\rho$,
and $\mathcal{M}^2_\eta$ by
\ba
\left( \begin{array}{c}
\varphi^+_0 \\ \varphi^+_1 \\ \varphi^+_2
\end{array} \right)
&=& \left( X_0, \ Y_1, \ Y_2 \right)
\left( \begin{array}{c} G^+ \\ S^+_1 \\ S^+_2 \end{array} \right),
\\
\left( \begin{array}{c}
\rho_0 \\ \rho_1 \\ \rho_2
\end{array} \right)
&=& \left( X_1, \ X_2, \ X_3 \right)
\left( \begin{array}{c} S^0_1 \\ S^0_2 \\ S^0_3 \end{array} \right),
\\
\left( \begin{array}{c}
\eta_0 \\ \eta_1 \\ \eta_2
\end{array} \right)
&=& \left( X_0, \ X_4, \ X_5 \right)
\left( \begin{array}{c} G^0 \\ S^0_4 \\ S^0_5 \end{array} \right),
\ea
respectively.
Thus,
\ba
\mathcal{M}_\varphi^2 Y_a = m_a^2 Y_a
& & \mbox{for} \ a = 1, 2,
\\
\mathcal{M}_\rho^2 X_b = \mu_b^2 X_b
& & \mbox{for} \ b = 1, 2, 3,
\\
\mathcal{M}_\eta^2 X_b = \mu_b^2 X_b
& & \mbox{for} \ b = 4, 5.
\ea
The $m_a^2$ ($a = 1, 2$) are the squared masses
of the charged scalars,
the $\mu_b^2$ ($b = 1, 2, 3$) are the squared masses
of the neutral scalars,
and the $\mu_b^2$ ($b = 4, 5$) are the squared masses
of the neutral pseudoscalars.
The decomposition of the Higgs doublets
in physical fields is given by
\be
\label{phi-dec}
\phi_j = 
\left( \begin{array}{c}
\left( X_0 \right)_j G^+ + \sum_{a=1}^2 \left( Y_a \right)_j S^+_a \\
2^{-1/2} \left[ v_j +
\sum_{b=1}^3 \left( X_b \right)_j S^0_b + i \left( X_0 \right)_j G^0
+ i \sum_{b=4}^5 \left( X_b \right)_j S^0_b
\right]
\end{array} \right).
\ee

\subsection{The light pseudoscalar}

The non-zero eigenvalues
of the mass matrix $\mathcal{M}_\eta^2$ of the pseudoscalars
are determined by
\be
\sigma \equiv \mu_4^2 + \mu_5^2
= \left( v_1^2 + v_2^2 \right) \left( \frac{A}{2} + k \aa \right)
\label{hdyur}
\ee
and 
\be
p \equiv \mu_4^2 \mu_5^2
= \aa \left( A v_1 v_2 - \aa v_0^2 \right) v^2,
\label{uidyr}
\ee
where we have defined
\be
\label{k}
k \equiv \frac{2 v_1 v_2}{v_1^2 + v_2^2}
= \frac{2 m_\mu m_\tau}{m_\mu^2 + m_\tau^2}
\approx \frac{1}{8.44}.
\ee

Fixing $S^0_4$ to be the lightest one of the two physical pseudoscalars,
\textit{i.e.}
\be
\begin{array}{rcl}
\mu_4^2 &=& \textstyle{\frac{1}{2}} \left( \sigma
- \sqrt{\sigma^2 - 4 p} \right),
\\*[1mm]
\mu_5^2 &=& \textstyle{\frac{1}{2}} \left( \sigma
+ \sqrt{\sigma^2 - 4 p} \right),
\end{array}
\ee
we see in equation~(\ref{uidyr}) that $\mu_4^2 = 0$
if either $\aa = 0$ or $\aa = A v_1 v_2 / v_0^2$; the latter case
means $\mm = 0$---see equation~(\ref{mm}).
This is easy to understand:
\begin{itemize}
\item If $\mu_\mathrm{m} = 0$,
then the $U(1)$ of~(\ref{U1}) is unbroken in the scalar potential.
This implies the existence of one physical neutral Goldstone boson,
corresponding to an extra
(\textit{i.e.}\ beyond $X_0$)
eigenvector $\left( 0, \, v_1, \, - v_2 \right)$ of $\mathcal{M}_\eta^2$
with eigenvalue $0$.
\item If $a_4 = 0$,
then there is an additional $U(1)$,
$\phi_0 \to e^{i \chi} \phi_0$,
unbroken in the scalar potential.
This implies the existence of one physical neutral Goldstone boson,
corresponding to an extra eigenvector
$\left( 1, \, 0, \, 0 \right)$ of $\mathcal{M}_\eta^2$
with eigenvalue $0$.
\item If both $\mu_\mathrm{m}$ and $a_4$ vanish,
then there are three $U(1)$ symmetries,
$\phi_j \to e^{i \alpha_j} \phi_j$ for $j = 1, 2, 3$,
unbroken in the scalar potential.
This might be thought to imply the existence
of two physical neutral Goldstone bosons.
However,
when both $\mu_\mathrm{m}$ and $a_4$ vanish,
$v_1$ also vanishes,\footnote{Equation~(\ref{mm})
holds trivially in this case.}
which means that one of those three $U(1)$ symmetries
is not spontaneously broken.
Therefore,
in that situation there is again \emph{one} physical neutral Goldstone boson,
corresponding to an extra eigenvector
$\left( v_2, \, 0, \, - v_0 \right)$ of $\mathcal{M}_\eta^2$
with eigenvalue $0$.
\end{itemize}

\begin{figure}[t]
\begin{center}
\epsfig{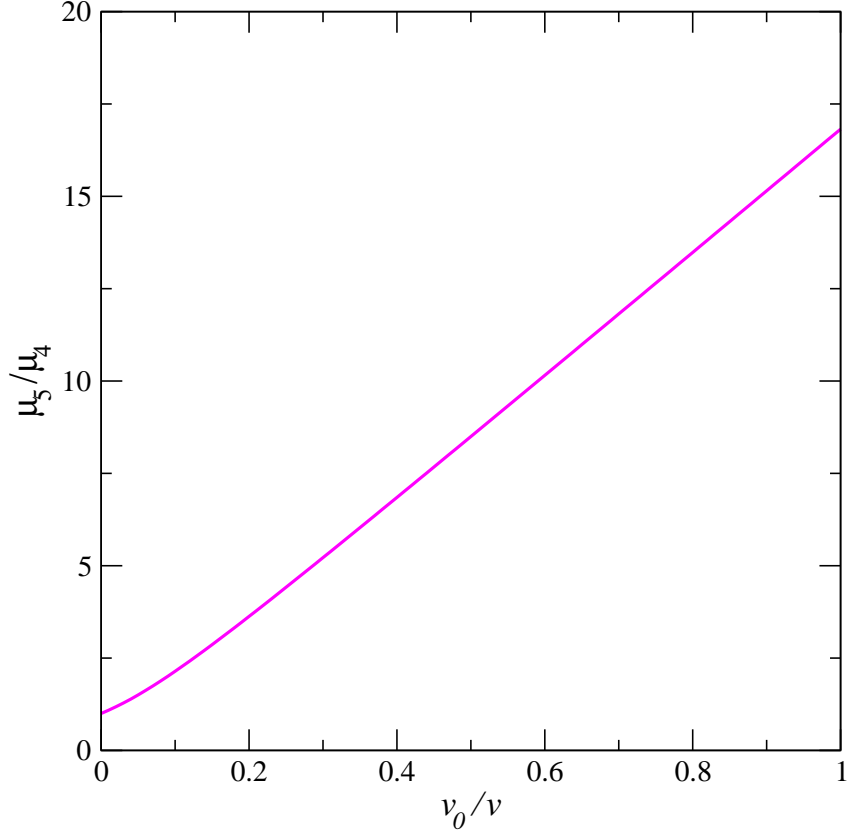}
\end{center}
\caption{The minimum possible value of $\mu_5 / \mu_4$
as a function of $v_0 / v$.
\label{chi2}}
\end{figure}

Since we know $v_1$ to be much smaller than $v_2$,
we expect,
in general,
both $\mm$ and $\aa$ to be relatively small,
and therefore we expect $S^0_4$ to be relatively light.
In order to quantify and qualify this expectation
we consider
\be
\label{jdtyr}
\frac{\mu_5^2}{\mu_4^2} = \frac{x + \sqrt{x^2 - 4}}{x - \sqrt{x^2 - 4}},
\ee
where $x \equiv \sigma \left/ \sqrt{p} \right.$.
Since
\be
\frac{\mathrm{d} \left( \mu_5^2 / \mu_4^2 \right)}{\mathrm{d} x} =
\frac{8}{\left( x - \sqrt{x^2 - 4}\, \right)^2 \sqrt{x^2 - 4}}
> 0,
\ee
$\mu_5^2 / \mu_4^2$ decreases when $x$ decreases.
Equations~(\ref{hdyur}) and (\ref{uidyr}) determine $x$ as a function
of $\aa$. Minimizing $x$ with respect to $\aa$, one finds 
\be
x_\mathrm{min} = 2 \sqrt{1 - r + \frac{r}{k^2}}\,, 
\quad \mbox{where} \ r \equiv \frac{v_0^2}{v^2}.
\ee
Inserting $x_\mathrm{min}$ into equation~(\ref{jdtyr}),
one obtains
\be
\left. \frac{\mu_5^2}{\mu_4^2} \right|_\mathrm{min} = \frac
{\sqrt{r - k^2 r + k^2} + \sqrt{r - k^2 r}}
{\sqrt{r - k^2 r + k^2} - \sqrt{r - k^2 r}}.
\ee
This lower bound on $\mu_5/\mu_4$ is depicted in figure~\ref{chi2}.
It is seen that,
unless $v_0$ is very small,\footnote{Note that
$\left. \mu_5/\mu_4 \right|_\mathrm{min} = 1$ for $v_0 = 0$.}
the lighter pseudoscalar will in general be ten or more times lighter
than the heavier pseudoscalar.
But,
$v_0$ cannot be too small,
lest the Yukawa coupling responsible for the top-quark mass
needs to be very large.

\subsection{The eigenvectors in the limit $v_1 = 0$}
\label{approximation}

As a preparation for the next section,
we now investigate the limit $v_1 = 0$ in detail.
In that limit $a_4$ and $\mu_\mathrm{m}$ must also vanish,
hence the scalar mass matrices are given by
\ba
\mathcal{M}^2_\varphi &=& \frac{1}{2} \left( \begin{array}{ccc}
- a_3 v_2^2 & 0 & a_3 v_0 v_2 \\
0 & - a_3 v_0^2 + \left( a_6 - 2 a_5 \right) v_2^2 & 0 \\
a_3 v_0 v_2 & 0 & - a_3 v_0^2
\end{array} \right),
\label{m+0} \\
\mathcal{M}^2_\rho &=& \frac{1}{2} \left( \begin{array}{ccc}
4 a_1 v_0^2 & 0 & 2 B v_0 v_2 \\
0 & A v_2^2 & 0 \\
2 B v_0 v_2 & 0 & 4 a_5 v_2^2
\end{array} \right),
\label{mr0} \\
\mathcal{M}^2_\eta &=& \frac{1}{2} \left( \begin{array}{ccc}
0 & 0 & 0 \\ 0 & A v_2^2 & 0 \\ 0 & 0 & 0
\end{array} \right).
\label{mi0}
\ea
From the positivity
of the mass matrices $\mathcal{M}^2_\varphi$ and $\mathcal{M}^2_\rho$
we then have
\be
a_3 < 0 \quad \mbox{and} \quad 4 a_1 a_5 > \left( a_2 + a_3 \right)^2,
\ee
respectively.
Notice that $a_1$ and $a_5$ must be positive,
even when $v_1 \neq 0$,
in order for the potential to have a minimum.

Equations~(\ref{m+0})--(\ref{mi0}) show that $\phi_1$ completely decouples
from $\phi_0$ and $\phi_2$ in the limit $v_1 = 0$.
In that limit one has
\be\label{limit}
m_2^2 = - \frac{a_3}{2}\, v_0^2 + \left( \frac{a_6}{2} - a_5 \right) v_2^2,
\quad
\mu_2^2 = \mu_5^2 = \frac{A}{2}\, v_2^2,
\quad
\mu_4^2 = 0,
\quad
Y_2 = X_2 = X_5 = \left( \begin{array}{c} 0 \\ 1 \\ 0 \end{array} \right).
\ee
Moreover,
equation~(\ref{m+0}) readily gives
\be\label{m1approx}
m_1^2 = - \frac{a_3}{2}\, v^2,
\quad
Y_1 =
\frac{1}{v} \left( \begin{array}{c} - v_2 \\ 0 \\ v_0 \end{array} \right).
\ee
while equation~(\ref{mr0}) leads to
\be
\mu^2_{1,3} = a_1 v_0^2 + a_5 v_2^2 \pm 
\sqrt{\left( a_1 v_0^2 - a_5 v_2^2 \right)^2
+ B^2 v_0^2 v_2^2},
\ee
and
\be
X_1 = \left( \begin{array}{c}
\cos \lambda \\ 0 \\ -\sin \lambda
\end{array} \right),
\quad
X_3 = \left( \begin{array}{c}
\sin \lambda \\ 0 \\ \cos \lambda
\end{array} \right),
\quad
\tan{2 \lambda} = \frac{B v_0 v_2}
{a_5 v_2^2 - a_1 v_0^2}.
\ee

\section{Phenomenology of the scalar sector}
\label{phenomenology}

The aim of this section is to demonstrate
that the model presented in this paper
complies with all the experimental constraints. 
We remind the reader
that the masses of the physical scalars,
and the mixing angles contained in the eigenvectors
$Y_a$ ($a = 1, 2$) and $X_b$ ($b = 1, \ldots, 5$),
are functions of the scalar-potential quartic couplings
$a_1, \dots, a_7$ and of $v_0 / v$.
It is beyond the scope of this paper
to perform a complete exploration of this large parameter space; 
we shall restrict ourselves to show
that it is possible to find a set of parameters such that
the scalar sector of the model
does not contradict any experimental results.
We will choose one such set and call it `the reference scenario'.

Although the present model is mainly designed for the lepton sector,
one may extend it to the quark sector, as mentioned in
section~\ref{potential}. 
The obvious way to do this is to stipulate that under
the $\zz$ symmetry in~(\ref{Z2})
the right-handed-quark gauge-$SU(2)$ singlets
transform with a minus sign.
Then,
only $\phi_0$ has Yukawa couplings to the quark sector.
In this way there are,
just as in the SM,
no flavour-changing neutral Yukawa interactions of the quarks.
This extension resembles in its spirit
a type-I two-Higgs-doublet model (2HDM). 
We require $v_0 \gtrsim 100$ GeV
in order to avoid a top-quark Yukawa coupling much larger than unity.

The Lagrangian for a generic multi-Higgs-doublet model (MHDM)
can be found in~\cite{GLOO07}.

\subsection{Constraints from $Z^0$ decay}

\paragraph{$Z^0$ decay into charged scalars}
In a MHDM,
the $Z^0$ couples to $S^+_a S^-_a$ with a universal strength,
independent of the details of charged-scalar mixing;
the relevant term in the Lagrangian is
\be
\frac{i g \left( s_W^2 - c_W^2 \right)}{2 c_W}\, Z_\mu
\sum_a
\left( S_a^+ \partial^\mu S_a^- - S_a^- \partial^\mu S_a^+ \right).
\ee
A model-independent lower bound on the masses $m_a$
of the charged scalars
can be derived from the invisible decay width of the $Z^0$.
Subtracting from it the SM decay width of the $Z^0$ into neutrinos,
the difference is compatible with zero,
leaving little room
for an additional decay of the $Z^0$ into charged scalars~\cite{DELPHI}.
This results in the bound~\cite{LEP} 
\be
\label{boundM+}
m_a > 43.7 \ \mbox{GeV} \ \left( 95\% \ \mbox{CL} \right),
\ a = 1, 2.
\ee

\paragraph{Higgs strahlung}
From LEP data,
a lower mass limit $m_h > 114.4$ GeV
has been deduced~\cite{RPP} for the SM Higgs particle $h$,
from the unobserved ``Higgs strahlung'' process
$e^+ e^- \to Z^\ast \to Z h$.
Note that this process is allowed only for scalars but not for
pseudoscalars~\cite{higgs-hunter}. 
In the present model,
all three scalars $S^0_{1,2,3}$
can in principle be produced by Higgs strahlung;
the relevant term in the Lagrangian is
\be
\frac{g m_Z}{2 c_W}\, Z_\mu Z^\mu \sum_{b=1}^3
\left( X_0 \cdot X_b \right) S^0_b,
\ee
where the quantity in parentheses denotes
the scalar products of the
vectors $X_0$ and $X_b$. 
In the limit $v_1 = 0$ the production of $S^0_2$ is suppressed
since $X_0 \cdot X_2 = 0$, as can be read off from equations~(\ref{X0})
and (\ref{limit}).
On the other hand, in that limit 
the strengths of the couplings of $S^0_1$ and $S^0_3$
are complementary, with 
$\left( X_0 \cdot X_1 \right)^2 + \left( X_0 \cdot X_3 \right)^2 = 1$.

\paragraph{Associated production}
The $Z^0$ can decay into a scalar--pseudoscalar pair~\cite{higgs-hunter};
the relevant term in the Lagrangian is
\be
\frac{g}{2 c_W}\, Z_\mu\,
\sum_{b=1}^3 \sum_{b^\prime = 4}^5
\left( X_b \cdot X_{b^\prime} \right)
\left( S_b^0 \partial^\mu S_{b^\prime}^0
- S_{b^\prime}^0 \partial^\mu S_b^0 \right).
\ee
The lightest pseudoscalar of our model,
$S^0_4$,
can in general be produced in this way
associated with either $S^0_1$ or $S^0_3$,
but not with $S^0_2$,
since $X_4 \cdot X_2 = 0$ in the limit of vanishing $v_1$.

\subsection{Constraints from other decays}

\paragraph{Decays of the charged scalars}
The 2HDM
has a single charged scalar $H^+$.
Assuming $\mbox{BR} \left( H^+ \to \tau^+ \nu_\tau \right)
+ \mbox{BR} \left( H^+ \to c \bar s \right) \simeq 1$,
the bound $m_{H^+} > 78.6$ GeV (95\% CL)
has been derived from the combined LEP data~\cite{RPP}. 
One cannot use this bound uncritically in the present model,
which has two charged scalars and in which
$\mbox{BR} \left( S_a^+ \to \mu^+ \nu_\mu \right)$
is certainly non-negligible.
Still,
the bound on $m_{H^+}$ suggests an estimate
of how much the bound~(\ref{boundM+}) can possibly be raised
by taking into account specific decay channels of $S_a^+$.

\paragraph{Other decays}
The transition $b \to s \gamma$ is important because it provides an indirect,
yet quite stringent,
lower bound on the charged-scalar
masses~\cite{gambino}.\footnote{See also~\cite{zochowski,kong}
and the references therein.} 
Since only $\phi_0$ has Yukawa couplings to the quarks
and the $S^+_2$ component of $\varphi^+_0$ is suppressed,
the lower bound from $b \to s \gamma$ applies only to $m_1$.
Vector mesons could possibly decay into a very light scalar
plus a photon~\cite{wilczek}, 
yielding a lower bound on the scalar mass.
This is relevant for the decay
$\Upsilon \left( 1S \right) \to S^0_4 \gamma$.
Loop corrections in the decay $Z^0 \to \bar b b$
are also important~\cite{denner}
in the 2HDM
for large $\tan \beta$. 
However,
since our model has features similar to a 2HDM
with $\tan\beta \sim 1$,
in which range this decay is not stringent~\cite{zochowski},
we will disconsider it in the following.

\subsection{``Safe'' scalar masses}

In the light of the above discussion we require
\be
\label{safe}
m_1 \gtrsim 350 \ \mbox{GeV}, \quad
\mu_1 \gtrsim 120 \ \mbox{GeV}, \quad
\mu_3 \gtrsim 120 \ \mbox{GeV}, \quad
\mu_4 > 10 \ \mbox{GeV}.
\ee
Some remarks are at order. 
In the 2HDM of type II the bound on the charged-scalar mass
from $b \to s \gamma$ is of the order of several hundred GeV,
much larger than the bound from direct LEP searches.
We have rather arbitrarily set that bound to 350 GeV in~(\ref{safe}),
by considering the results obtained in~\cite{gambino}
for $\tan \beta \sim 1$
and taking into account the considerable uncertainty
in the computation of the corresponding $B$-meson decay.
The bounds on $\mu_1$ and $\mu_3$ have been stipulated
in order to definitely avoid production via Higgs strahlung.
Finally,
the lower bound on $\mu_4$ stems from the wish to avoid
any problems from $\Upsilon(1S) \to S^0_4 \gamma$.
We have not put lower bounds on $m_2$,
$\mu_2$,
and $\mu_5$
in~(\ref{safe}) because, from the discussions in the previous paragraphs, 
we conclude that there are no
really stringent bounds on these masses.
Of course,
these masses should not be too small.
In any case, numerically it will turn out that
if we fulfill the constraints of~(\ref{safe}),
then also $m_2$ and $\mu_2$ will be reasonably large. 
Moreover,
we bear in mind that in our model $\mu_5 \gg \mu_4$ holds anyway.

\subsection{A reference scenario}

In table~\ref{table} we have written down a set of values for the eight
parameters of the model,
which we define to be our `reference scenario'.
All input values are of order one,
except $a_3$ which is somewhat
larger because it is responsible
for a large mass $m_1$---see equation~(\ref{m1approx}).
In table~\ref{table}, $|a_4|_\mathrm{max} = A v_1 v_2/v_0^2$ is the
maximal value of $|a_4|$, obtained from equation~(\ref{mm}).
\begin{table}
\begin{center}
\begin{tabular}{|cccccccc|}\hline
$v_0/v$ & $a_1$ & $a_2$ & $a_3$ & $\aa/\aa_\mathrm{max}$ & 
$a_5$ & $a_6$ & $a_7$ \\ \hline\hline
$1/\sqrt{2}$ &
2.5 & 3 & $-5$ & 0.4 & 1.5 & 2 & 3 \\ \hline
\end{tabular}
\end{center}
\caption{Input values for the reference scenario.\label{table}}
\end{table}

Taking the input from table~\ref{table}
and performing  a numerical calculation,
we obtain 
\be
\label{utrkd}
\begin{array}{lclclclclcl}
m_1 &=& 389.2 \ \mbox{GeV}, & & m_2 &=& 245.8 \ \mbox{GeV},
\\
\mu_1 &=& 434.8 \ \mbox{GeV}, & &
\mu_2 &=& 171.9 \ \mbox{GeV}, & &
\mu_3 &=& 231.8 \ \mbox{GeV},
\\
\mu_4 &=& 14.3 \ \mbox{GeV}, & &
\mu_5 &=& 173.8 \ \mbox{GeV}. & & & 
\end{array}
\ee
These values agree well
with the ones computed
from the approximate formulae of section~\ref{approximation}. 
For instance, $\mu_2 \simeq \mu_5$ in~(\ref{utrkd}).
The masses~(\ref{utrkd})
satisfy
the conditions~(\ref{safe}) for ``safe'' masses.

Next we check the reference scenario
against electroweak precision data
by using the oblique parameters~\cite{STU} $S$,
$T$,
and $U$. 
For a MHDM
we take the formula for $T$ in~\cite{GLOO07}
(for computations of $T$ in the 2HDM,
see e.g.~\cite{higgs-hunter,bertolini,rho}), 
which gives,
when applied to the present model
\ba
\label{T}
T &=& \frac{1}{16 \pi s_W^2 m_W^2} \left\{
\sum_{a=1}^2 \sum_{b=1}^5 \left( Y_a \cdot X_b \right)^2 
F \left( m_a^2, \, \mu_b^2 \right) \right.
\no & &
- \sum_{b=1}^3 \sum_{b^\prime = 4}^5
\left( X_b \cdot X_{b'} \right)^2 
F \left( \mu_b^2, \, \mu_{b^\prime}^2 \right)
\no & &
+ 3 \sum_{b=1}^3 \left( X_0 \cdot X_b \right)^2 \left[ 
F \left( m_Z^2, \, \mu_b^2 \right) - F \left( m_W^2, \, \mu_b^2 \right)
\right]
\no & & \left.
- 3 \left[ F \left( m_Z^2, \, m_h^2 \right) - F\left( m_W^2, \, m_h^2 \right)
\right]
\vphantom{\sum_{a=1}^2 \sum_{b=1}^5} \right\},
\ea
where
\be
F \left( x, \, y \right) = \frac{x + y}{2} - \frac{x y}{x - y}\,
\ln{\frac{x}{y}}.
\ee
In equation~(\ref{T}),
$m_W$ and $m_Z$ are the masses of the $W^\pm$ and $Z^0$ gauge bosons,
respectively,
$s_W^2 = 1 - m_W^2/m_Z^2$,
and $m_h$ is the mass of the SM Higgs boson.
One may also write down formulae for $S$ and for $U$
by applying the results in~\cite{GLOO08}.
Taking the central value $m_h = 87$ GeV from recent SM fits~\cite{mh}
and using the scalar masses and diagonalizing matrices
of the reference scenario,
we have obtained $S = 0.046$,
$T = -0.162$,
and $U = -0.002$.
These values are compatible with the fit results for the oblique parameters
given in~\cite{erler}.
We note that,
while all the individual contributions to $S$ and to $U$ are small
and no excessive cancellations occur among them,
this is not so for $T$:
considering separately the first
and the second terms in the right-hand side (RHS) of equation~(\ref{T}),
each of them is one order of magnitude larger
than the final result $T=-0.162$;
however,
those two contributions have opposite signs
(note that $F \geq 0$),
leading to a partial cancellation.
Nevertheless,
a certain amount of tuning of the input parameters is expedient 
to achieve the correct order of magnitude of $T$,
as will be discussed in the next paragraph. 
The numerical value of third term of the RHS of equation~(\ref{T})
is naturally one order of magnitude smaller
than the values of the first and second terms,
and the SM subtraction in the forth term is numerically a tiny effect. 

\begin{figure}
\begin{center}
\epsfig{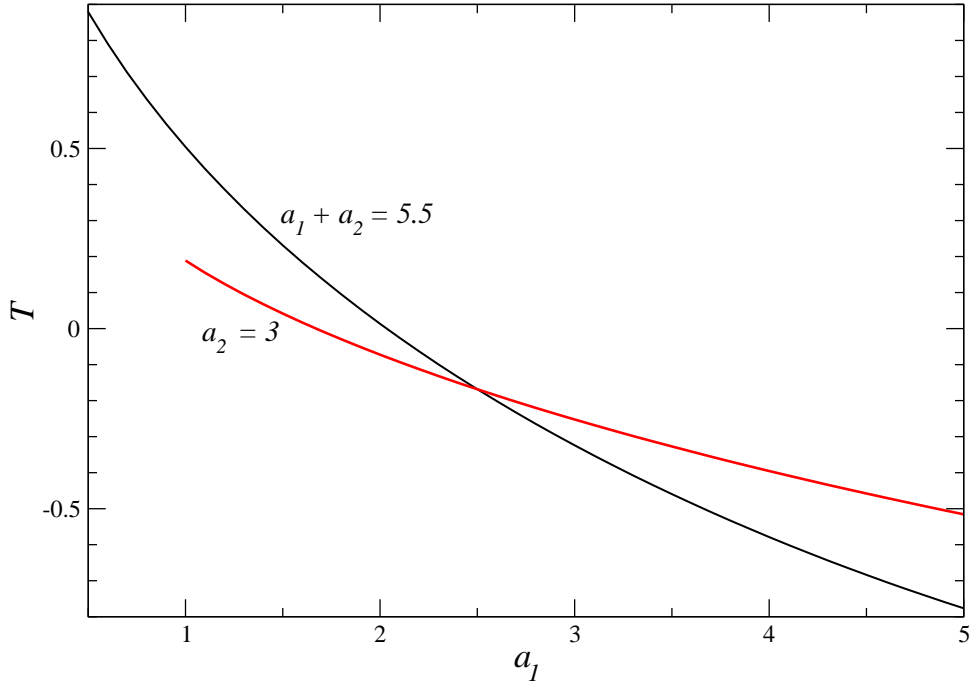}
\end{center}
\caption{The oblique parameter $T$ as a function of $a_1$. 
The input parameters not shown in the plot are as in table~\ref{table}. 
\label{tfig}}
\end{figure}

In figure~\ref{tfig}
we have attempted to illustrate
the dependence of $T$ on the input parameters $a_1$ and $a_2$;
these occur only in the mass matrix of the neutral scalars~(\ref{MR})
and might,
therefore,
be able to disturb the cancellation
between the first and second terms in the RHS in equation~(\ref{T}).
In figure~\ref{tfig} we have plotted two curves.
In the first one we have fixed $a_2 = 3$
to its value in the reference scenario,
whereas in the other one we have fixed $a_1 + a_2 = 5.5$;
the crossing point of the curves corresponds to the reference scenario. 
The figure illustrates nicely the tuning
required for keeping $T$ small.
We note that the curve with fixed $a_2$ begins at $a_1 = 1$,
where $\mu_3= 115$ GeV is outside the region of ``safe'' masses,
but $\mu_3$ quickly grows with $a_1$.

\section{Conclusions}
\label{concl}

In this paper we have presented a model
for the lepton sector
based on the family symmetry $O(2)$.
The model has an obvious extension to the quark sector,
by coupling to the quarks
only the Higgs doublet $\phi_0$ which transforms trivially under $O(2)$.
The smallness of the masses of the light neutrinos
is explained in our model through the seesaw mechanism.
The reflection symmetry contained in $O(2)$
acts as a $\mu$--$\tau$ interchange symmetry,\footnote{In appendix~C
we show that it is possible to replace the reflection symmetry
by a non-standard $CP$ transformation;
in that version of the model there is no $O(2)$ family symmetry.}
which---together with the $U(1) \subset O(2)$---enforces
diagonal Yukawa-coupling matrices
and a neutrino mass matrix of the form~(\ref{Mnu}).
Consequently,
the model predicts $\theta_{23} = \pi/4$ and $\theta_{13} = 0$
at the tree level.
In that respect the model in this paper
is practically identical to the one in~\cite{z2};
in both models the lepton flavour violation
resides only in the mass matrix of the right-handed neutrinos.
It has been shown~\cite{soft}
that models of this class
are safe from flavour-changing neutral interactions.

The crucial difference between the present model
and the one in~\cite{z2} is that,
in the $O(2)$ model,
we allow a non-trivial transformation
of the Higgs doublets $\phi_1$ and $\phi_2$
under the $U(1) \subset O(2)$.
In this way,
we obtain a relation
between the smallness of the mass ratio $m_\mu/m_\tau$
and the small mass $\mu_4$ of one of the two pseudoscalars of the model;
indeed,
that pseudoscalar is almost a Goldstone boson
and only a soft $U(1)$ breaking in the scalar potential prevents $\mu_4$
from being exactly 
zero. 
On the other hand,
that soft breaking must necessarily be small
in order to reproduce the small value of $m_\mu/m_\tau = v_1/v_2$,
determined by the ratio of the VEVs of $\phi_1$ and 
$\phi_2$.\footnote{It is interesting to note that, if both the $U(1)$
  and the reflection symmetry $s$ are softly broken in the scalar
  potential, then $v_1 = 0$ still implies $\mu_m = 0$, $a_4 = 0$,
  and a Goldstone boson. Thus, the prediction $\mu_4 \ll \mu_5$
  remains unaltered.}

It had previously been realized~\cite{zochowski,krawczyk} that,
in the 2HDM,
one of the neutral scalars could be quite light
without contradicting any experimental constraints.
We have attempted
to show that the same holds in our three-Higgs-doublet model.
Actually,
our model not only predicts the light pseudoscalar, 
it also predicts the near equality
of the mass of one of the scalars
and the mass of the heaviest pseudoscalar,
and, in addition, specific features in scalar mixing,
resulting from the near decoupling of $\phi_1$
from $\phi_0$ and $\phi_2$,
due to the smallness of $m_\mu/m_\tau$. 
We have thus demonstrated that,
within our $O(2)$ model,
the connection between the lepton and scalar sectors
can be much tighter than usually thought of.

\appendix

\setcounter{equation}{0}
\renewcommand{\theequation}{A\arabic{equation}}

\section{The group $O(2)$}

\paragraph{Definition and characterization}
$O(2)$ is the group of rotations and reflections of the plane. 
It is generated by rotations $g \left( \theta \right)$,
with angle $\theta$,
around the center of the coordinate system,
and by the reflection $s$ about the $x$-axis.
Allowing the angle $\theta$ to vary over $\mathbbm{R}$,
the properties of these group elements,
which fully characterize the group,
are 
\be
g \left( \theta + 2\pi \right) = g \left( \theta \right),
\quad 
g \left( \theta_1 \right) g \left( \theta_2 \right)
= g \left( \theta_1 + \theta_2 \right),
\quad
s^2 = e,
\quad 
s\, g \left( \theta \right) s = g \left( -\theta \right).
\ee

\paragraph{Irreducible representations}
There are two singlet irreducible representations of $O(2)$:
\be
\underline{1}: \ g \left( \theta \right) \to 1, \ s \to 1
\quad \mbox{and} \quad
\underline{1}^\prime: \ g \left( \theta \right) \to 1, \ s \to -1.
\ee
Furthermore,
$O(2)$ has a countably infinite set
of doublet irreducible representations,
numbered by $n \in \mathbbm{N}$:
\be
\label{2n}
\underline{2}^{(n)}: \ g \left( \theta \right) \to 
\left( \begin{array}{cc}
e^{i n \theta} & 0 \\ 0 & e^{-i n \theta}
\end{array} \right), \
s \to \left( \begin{array}{cc} 0 & 1 \\ 1 & 0  \end{array} \right).
\ee

\paragraph{Tensor product $\underline{2}^{(m)} \otimes \underline{2}^{(n)}$}
We assume that the matrices in~(\ref{2n})
act on an orthonormal basis $\{ e_1,\, e_2 \}$.
In the product $\underline{2}^{(m)} \otimes \underline{2}^{(n)}$
we must distinguish two cases.
If $m>n$,
then
\be
\label{uftyp}
\underline{2}^{(m)} \otimes \underline{2}^{(n)}
=
\underline{2}^{(m+n)} \oplus \underline{2}^{(m-n)}.
\ee
The irreducible representations in the right-hand side of~(\ref{uftyp})
have basis vectors
\be
\underline{2}^{(m+n)}: \
e_1 \otimes e_1, \ e_2 \otimes e_2
\quad \mbox{and} \quad
\underline{2}^{(m-n)}: \
e_1 \otimes e_2, \ e_2 \otimes e_1.
\ee
If $m=n$,
then
\be
\label{utste}
\underline{2}^{(n)} \otimes \underline{2}^{(n)}
=
\underline{1} \oplus \underline{1}^\prime \oplus \underline{2}^{(2n)}.
\ee
The irreducible representations in the right-hand side of~(\ref{utste})
have basis vectors
\be
\underline{1}: \
\frac{1}{\sqrt{2}} \left( e_1 \otimes e_2 + e_2 \otimes e_1 \right),
\quad
\underline{1}^\prime: \
\frac{1}{\sqrt{2}} \left( e_1 \otimes e_2 - e_2 \otimes e_1 \right),
\quad
\underline{2}^{(2n)}: \
e_1 \otimes e_1, \ e_2 \otimes e_2.
\ee

\setcounter{equation}{0}
\renewcommand{\theequation}{B\arabic{equation}}

\section{Comparison of the present model
with the model of softly broken lepton numbers}
\label{comparison}

\paragraph{The $\zz$ model}
The model presented in this paper---let us call it
``$O(2)$ model''---is quite similar to the model
proposed by two of us a few years ago~\cite{z2}---let us call it
``$\zz$ model''.
The $\zz$ model has the same fermion and scalar multiplets
as the $O(2)$ model.
Both the $\zz$ and $O(2)$ models
have the $s$ of~(\ref{s}) and the $\zz$ of~(\ref{Z2})
as symmetries.
However,
instead of the $U(1)$ of~(\ref{U1}),
employed as a symmetry in the $O(2)$ model,
the $\zz$ model requires the conservation,
in all terms of dimension four in the Lagrangian,
of the three family lepton numbers.
As a consequence,
the Yukawa Lagrangian of the $\zz$ model has,
beyond the terms in equation~(\ref{LY}),
one further term:
\be
\label{y5}
\mathcal{L}_Y = \cdots
- y_5 \left( \bar D_{\mu  L} \phi_2 \mu_R 
+ \bar D_{\tau  L} \phi_1 \tau_R \right)
+ \mbox{H.c.}
\ee
Therefore,
in the $\zz$ model the ratio between the muon and tau masses is
\be
\frac{m_\mu}{m_\tau} = \left|
\frac{y_4 v_1 + y_5 v_2}{y_4 v_2 + y_5 v_1} \right|.
\ee

\paragraph{Symmetry group $O(2)$ in the $\zz$ model}
It was noted as a side remark in~\cite{su5}
that the $\zz$ model also has family symmetry $O(2)$.
This group $O(2)$ is generated by
the $\mu$--$\tau$ interchange symmetry $s$
together with
the $U(1)$ of the lepton number $L_\mu - L_\tau$.
Replacing $\phi_1$ and $\phi_2$
by $\phi_\pm \equiv \left( \phi_1 \pm \phi_2 \right)
\left/ \sqrt{2} \right.$,
we see that,
under that $O(2)$,
$\phi_+$ transforms as a $\underline{1}$
and $\phi_-$ as a $\underline{1}^\prime$.
The $O(2)$ model,
on the other hand,
has two Higgs doublets transforming as a $\underline{2}^{(2)}$ of $O(2)$,
instead of as a $\underline{1} \oplus \underline{1}^\prime$;
one further difference is that the $U(1)$ group in the $O(2)$ model
is not really $L_\mu - L_\tau$,
\textit{cf}.~(\ref{U1}).

\paragraph{Naturally small $m_\mu / m_\tau$ in the $\zz$ model}
In~\cite{smallratio} an additional symmetry,
dubbed $K$,
was introduced into the $\zz$ model
in order to provide a technically natural explanation
for the smallness of $m_\mu / m_\tau$.
Under $K$,
$\phi_1$ and $\mu_R$ change sign
while all other fields remain invariant.
The symmetry $K$ eliminates
the $y_5$ term---see equation~(\ref{y5})---from the Yukawa Lagrangian
of the $\zz$ model,
thus obtaining $m_\mu / m_\tau = \left| v_1 / v_2 \right|$
just as in the $O(2)$ model.
We want to stress that,
from the point of view of neutrino masses and lepton mixing,
the $O(2)$ model of the present paper
is equivalent to the $\zz$ model of~\cite{z2}
and also to the $\zz$ model with the additional symmetry $K$
of~\cite{smallratio}.
The difference lies in the scalar potential,
which in the $O(2)$ model is both different and more restricted.
Indeed,
in the $\zz$ model with a softly broken symmetry $K$,
the $a_4$ term is absent from the scalar potential;
on the other hand,
there are extra terms
\be
V = \cdots
+ b_1 \left[ \left( \phi_1^\dagger \phi_2 \right)^2
+ \left( \phi_2^\dagger \phi_1 \right)^2 \right]
+ \left\{ b_2 \left[ \left( \phi_0^\dagger \phi_1 \right)^2
+ \left( \phi_0^\dagger \phi_2 \right)^2 \right]
+ \mbox{H.c.} \right\},
\ee
with $b_1$ real but $b_2$ in general complex.
The model of~\cite{smallratio} has the advantage,
over the $O(2)$ model,
that $m_\mu / m_\tau$ is small in a technically natural sense;
indeed,
in that model $v_1 \neq 0$ only obtains when $K$ is softly broken
by the $\mu_3$ term,
while in the $O(2)$ model $v_1 \neq 0$,
even if $\mu_3 = 0$,
because of the $a_4$ term.
The advantage of the $O(2)$ model is its prediction
of a light pseudoscalar---a prediction
inexistent in the model of~\cite{smallratio}.

\setcounter{equation}{0}
\renewcommand{\theequation}{C\arabic{equation}}

\section{Substitution of the symmetry $s$
by a non-diagonal $CP$ symmetry}
\label{CP}

In the model suggested in this paper
it is possible to use,
instead of the $\mu$--$\tau$ interchange symmetry $s$,
the non-trivial $CP$ symmetry~\cite{CP,app}
\be
CP: \left\{ \begin{array}{rcl}
D_{\alpha L} &\to& i S_{\alpha \beta} \gamma^0 C \bar D_{\beta L}^T,
\\
\alpha_R &\to& i S_{\alpha \beta} \gamma^0 C \bar \beta_R^T,
\\
\nu_{\alpha R} &\to& i S_{\alpha \beta} \gamma^0 C \bar \nu_{\beta R}^T,
\\
\phi_0 &\to& \phi_0^\ast,
\\
\phi_1 &\to& \phi_2^\ast,
\\
\phi_2 &\to& \phi_1^\ast,
\end{array} \right.
\quad
\mbox{where} \
S = \left( \begin{array}{ccc}
1 & 0 & 0 \\ 0 & 0 & 1 \\ 0 & 1 & 0
\end{array} \right)
\ee
and $C$ is the Dirac--Pauli charge conjugation matrix.
This $CP$ symmetry commutes with both the $U(1)$ of~(\ref{U1})
and the $\zz$ of~(\ref{Z2}),
so that,
in this case,
the model has symmetry $CP \times U(1) \times \zz$
instead of $O(2) \times \zz$.
Instead of equation~(\ref{LY}) we would then have
\ba
\mathcal{L}_Y & = &
- y_1\, \bar D_{eL} \tilde \phi_0 \nu_{eR}
- \left( y_2\, \bar D_{\mu L} \tilde \phi_0 \nu_{\mu  R}
+ y_2^\ast\, \bar D_{\tau L} \tilde \phi_0 \nu_{\tau R} \right)
\no &&
- y_3\, \bar D_{e L} \phi_0 e_R
- \left( y_4\, \bar D_{\mu  L} \phi_1 \mu_R 
+ y_4^\ast\, \bar D_{\tau  L} \phi_2 \tau_R \right)
+ \mbox{H.c.,}
\label{LY2}
\ea
with real $y_{1,3}$.
We would end up with~\cite{valle}
\be
\label{mnu2}
\mnu = \left( \begin{array}{ccc}
x & y & y^\ast \\ y & z & w \\ y^\ast & w & z^\ast
\end{array} \right)
\ee
$x$ and $w$ being real.
Such a model predicts~\cite{CP} maximal atmospheric-neutrino mixing
($\theta_{23} = \pi / 4$) but,
instead of $U_{e3} = 0$,
it predicts~\cite{HS,CP} 
$\left| U_{\mu i} \right| = \left| U_{\tau i} \right|$
for all $i = 1, 2, 3$
($U$ is the lepton mixing matrix),
which leads to $\sin{\theta_{13}} \cos{\delta} = 0$, with $\delta$ being
the $CP$-violating phase in the mixing matrix.
Although this condition permits $\theta_{13} = 0$,
it can be shown that the more general case 
is that of maximal $CP$ violation~\cite{CP}
\textit{i.e.}~$\delta = \pm \pi/2$.
The scalar potential is the same as in equation~(\ref{pot})
with the proviso~(\ref{transf}).

\end{document}